\documentclass[rnote,traditabstract]{aa} 
\usepackage{graphicx}
\usepackage{txfonts}
\begin{document}
\title{Search for the shortest variability at gamma rays\\in flat-spectrum radio quasars}

\author{L. Foschini\inst{1}, 
			G. Ghisellini\inst{1},
			F. Tavecchio\inst{1},
			G. Bonnoli\inst{1},
			A. Stamerra\inst{2,3}
}

\institute{
INAF - Osservatorio Astronomico di Brera, via E. Bianchi 46, 23807, Merate (LC), Italy\\
\email{luigi.foschini@brera.inaf.it}
   \and
Universit\`a di Siena, via Roma 56, 53100, Siena, Italy
	\and
INFN Pisa, Via Buonarroti 2, 56127, Pisa, Italy\\
             }

   \date{Received --; accepted --}
 
\abstract{
We report about the search for short-term variability in the high-energy $\gamma$-ray energy band of three flat-spectrum radio quasars (3C~454.3, 3C~273, PKS~B1222$+$216), whose flux at $E>100$~MeV exceeded the value of $10^{-5}$~ph~cm$^{-2}$~s$^{-1}$ for at least one day. Although, the statistics was not yet sufficient to effectively measure the characteristic time scale, it allowed us to set tight upper limits on the observed doubling time scale ($<2-3$~hours) -- the smallest measured to date at MeV energies --, which can constrain the size of the $\gamma$-ray emitting region. The results obtained in the present work favor the hypothesis that $\gamma$ rays are generated inside the broad-line region.}


\authorrunning{L. Foschini et al.}
\titlerunning{Gamma-ray Fast Variability in FSRQ}

\maketitle

\section{Introduction}
The latest generation of ground-based Cerenkov telescope, like HESS and MAGIC, has detected flux changes on a few minutes time scale in the BL Lac Objects PKS 2155$-$304 (Aharonian et al. 2007) and Mkn 501 (Albert et al. 2007). These episodes, although exceptional, have severely challenged the common assumptions about the size and location of the emitting region. According to these models, the emitting region of size $r>r_{\rm g}=GM/c^2$ (being $r_{\rm g}$ the gravitational radius of the central black hole with mass $M$) is located at a distance $R \sim r/\psi$ from the central spacetime singularity, where $\psi$ is the aperture of the jet (generally $\psi \sim 0.1-0.25$, see Ghisellini \& Tavecchio 2009, Dermer et al. 2009). Changes in the emitted radiation from this blob occur on a characteristic observed time $\tau/(1+z)>r/c\delta$, where $\delta$ is the Doppler factor of the jet. 

It is worth noting that the dissipation zone should be sufficiently far from the central black hole to avoid pair production due to the interaction with the photons emitted mainly by the accretion disc and the broad-line region (BLR). BL Lac Objects are known to have a photon-starved environment: the accretion disc in these sources is weak, likely an advection-dominated disc (ADAF), which is also inefficient in ionizing the BLR. Therefore, the variability of a few minutes observed in these BL Lac Objects can be explained by assuming very small blobs at small $R$: the paucity of soft photons makes it possible for GeV photons to escape without pair production. 

This is no more possible in the case of FSRQs, where discs with accretion rates of a few tens of percent of the Eddington rate are present, which in turn efficiently photoionize the BLR, generating broad and strong optical emission lines. This rich environment (disc, BLR) makes the cooling of relativistic electrons more efficient through the external Compton (EC) processes, but, on the other hand, if the emitting blob is too close to the accretion disc to be sufficiently small for the short variability, then the pair production could severely absorb the GeV photons. 

Some researchers have proposed different solutions to this problem, which are basically relying on the increase of the Doppler boosting or some special jet structure (Begelman et al. 2008, Finke et al. 2008, Ghisellini \& Tavecchio 2008, Giannios et al. 2009). The latter, based on the hypothesis that there are very compact relativistic blobs within the jet (``needle/jet'' model, Ghisellini \& Tavecchio 2008; ``jet-in-jet'' model, Giannios et al. 2009), can be adapted to both BL Lac Objects and FSRQs, although it seems an \emph{ad hoc} solution at the present stage and, perhaps, it is necessary to better assess this type of solution by comparing with more and more observations. Therefore, the most critical aspect of these theories would be the discovery of variability of a few minutes in the case of FSRQs.  

The shortest time scales measured in FSRQs is of about half an hour, as observed in the hard X-ray emission (20$-$40~keV) of NRAO~530 in 2004 February (Foschini et al. 2006). However, given the energy range, it would be possible that this short flare could have been generated by the X-ray corona, although the emitted power ($\sim 8\times 10^{47}$~erg~s$^{-1}$) seems a bit large to be due to an unbeamed component. In addition, an increase of the radio polarization around the time of the flare suggests that it was indeed related to the jet, although the coarse observation sequence did not allow to set tight constraints and doubts shadow the conclusions.

Anyway, if short time scales could be observed in FSRQs at hard X-rays, which are sampling low energy part of the inverse-Compton hump of the spectral energy distribution (SED), then there is the possibility that similar variability is present also in the high-energy part ($E > 100$~MeV). At these energies, there is no more doubt that the emission is coming from the jet, since no other structures are known to generate such energetic photons at cosmological distances. 

The known flux variations at high-energy $\gamma$ rays ($E>100$~MeV) in FSRQs are generally in the range of a few hours, in agreement with the current paradigm.  The best examples measured by \emph{CGRO}/EGRET during nineties are PKS~1622$-$29 ($\sim 4$ hours, Mattox et al. 1997) and 3C~279 ($\sim 8$ hours, Wherle et al. 1998). The early results from \emph{Fermi} satellite obtained in the latest years have confirmed such time scales: PKS~1454$-$354 (Abdo et al. 2009), PKS~1502$+$106 (Abdo et al. 2010a), PKS~B1510$-$089 (Tavecchio et al. 2010), 3C~454.3 (Foschini et al. 2010, Tavecchio et al. 2010, Ackermann et al. 2010) and 3C~273 (Abdo et al. 2010b). 

The availability of the \emph{Fermi} satellite provides a plethora of $\gamma$-ray data where to search for short variability events. The Large Area Telescope (LAT, Atwood et al. 2009) onboard \emph{Fermi} represents the state of the art of $\gamma$-ray space instruments, with an increase of sensitivity of a factor $\sim 20-30$ with respect to its predecessors. \emph{Fermi} operates in scanning mode, i.e. it performs an all-sky survey every 3 hours (two orbits). This is the first continuous monitoring of the high-energy $\gamma$-ray sky and offers the unique possibility to study the blazar population, the duty cycle of individual sources, and to catch the most powerful outbursts. The bad side of the thing is that the scanning mode hampers the probing of subhour variability, because the source is not always at the LAT's boresight, where the instrument has its best performance. A tentative to get over this obstacle out has been done in 2010 April, when 3C~454.3 underwent an intense ourburst with flux above 100 MeV in excess of $10^{-5}$~ph~cm$^{-2}$~s$^{-1}$. During this episode, \emph{Fermi}/LAT performed a pointed observation staring at the FSRQ (2010 April 5$-$8). The latter, however, did not collaborate and remained almost constant with poorly significant variations ($\sim 3$~hours) and began to be variable only toward the end of the special observation (Foschini et al. 2010, Ackermann et al. 2010). 

Therefore, we decided to expand our search to other -- possibly more active -- periods and sources. We searched for other cases of FSRQ with high-energy $\gamma$-ray flux exceeding $10^{-5}$~ph~cm$^{-2}$~s$^{-1}$ for at least one day, in order to have the best available statistics. We found three FSRQs fulfilling this criterion, which were 3C~454.3 ($z=0.859$), 3C~273 ($z=0.158$), and PKS~B1222$+$216 ($z=0.432$). We report here the results of this analysis. There were two more cases that could have been of interest, but they were discarded. PKS~B1510$-$089 and PKS~1830$-$211 have exceeded the threshold flux for a few hours (Ciprini et al. 2010, Tavecchio et al. 2010), but not for at least one day, and therefore we did not consider them. Moreover, at the time of writing this work (Christmas 2010), 3C~454.3 exceeded again the threshold, but we limited our data set at the end of 2010 November.

\begin{figure}
\centering
\includegraphics[angle=270,scale=0.35]{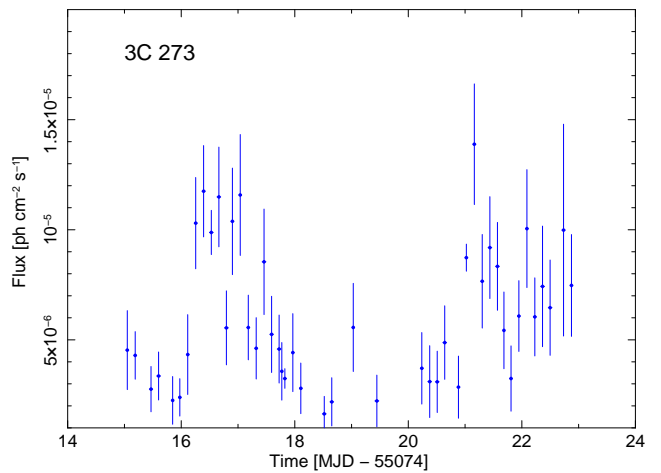}
\caption{Light curves of 3C 273 ($E>100$~MeV) between 2009 September 15 and 23. Time starts on 2009 August 31 (MJD 55074), so to have in the abscissa the days of September. Time bins are of the order of a few thousands of seconds and are too small to be visible.}
\label{fig:curva3c273}
\end{figure}

\begin{figure*}
\centering
\includegraphics[angle=270,scale=0.35]{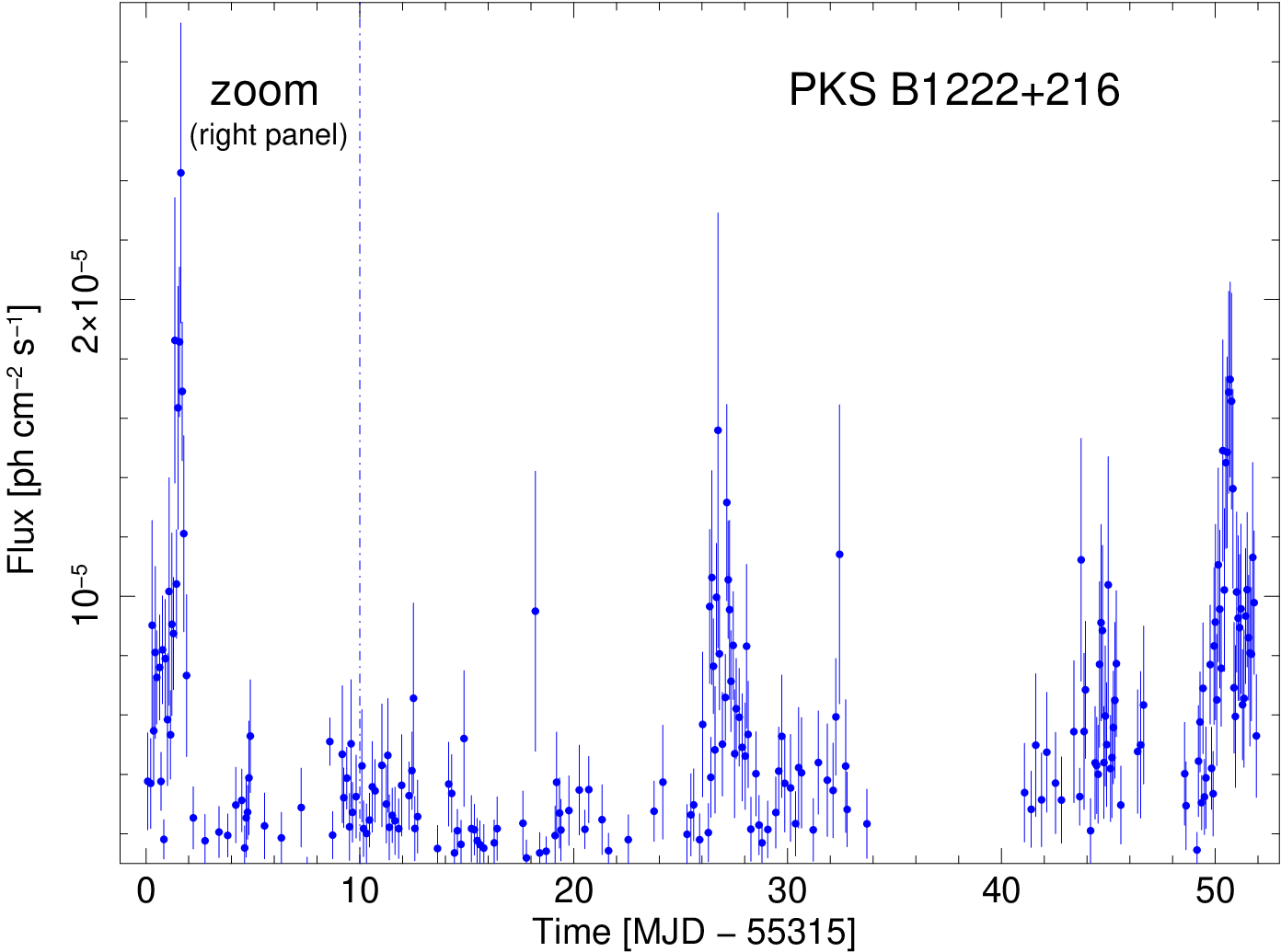}
\includegraphics[angle=270,scale=0.35]{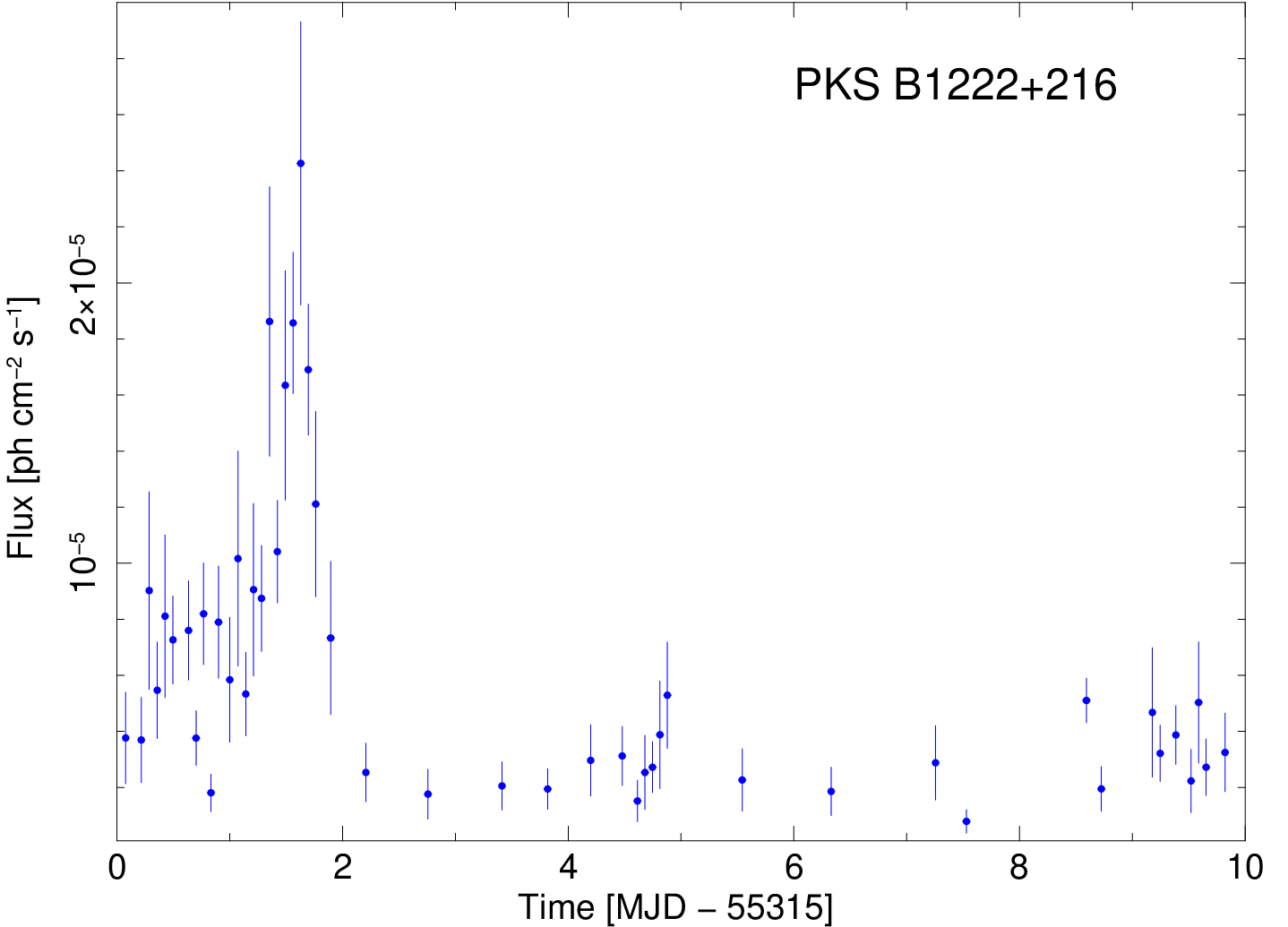}
\caption{Light curves of PKS~B1222$+$216 ($E>100$~MeV). (\emph{left panel}) Light curve in the period 2010 April 29 (MJD 55315) and June 20. (\emph{right panel}) zoom of first ten days of the curve. Time bins are of the order of a few thousands of seconds and are too small to be visible.}
\label{fig:curva1222}
\end{figure*}

\begin{figure*}
\centering
\includegraphics[angle=270,scale=0.35]{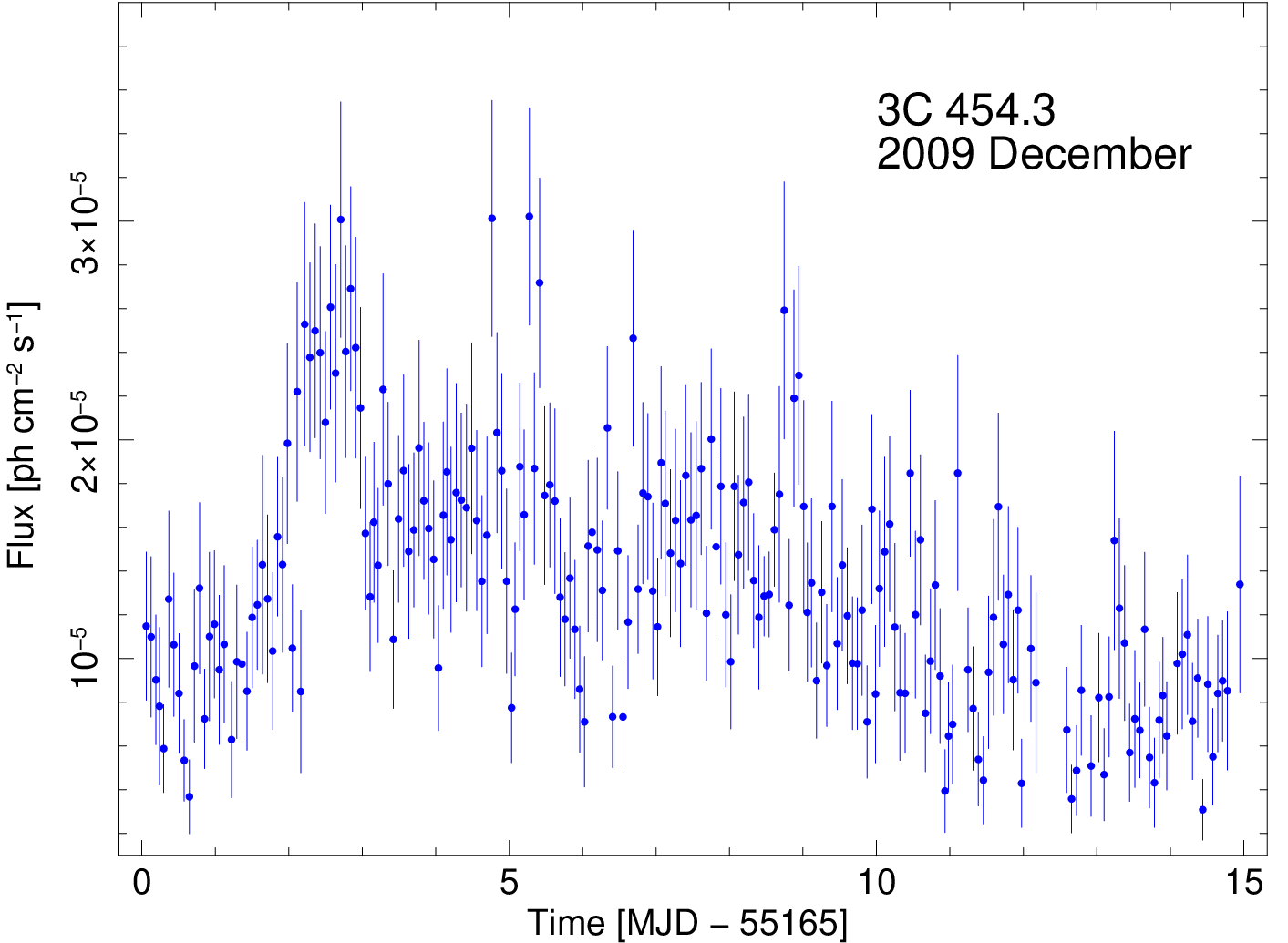}
\includegraphics[angle=270,scale=0.35]{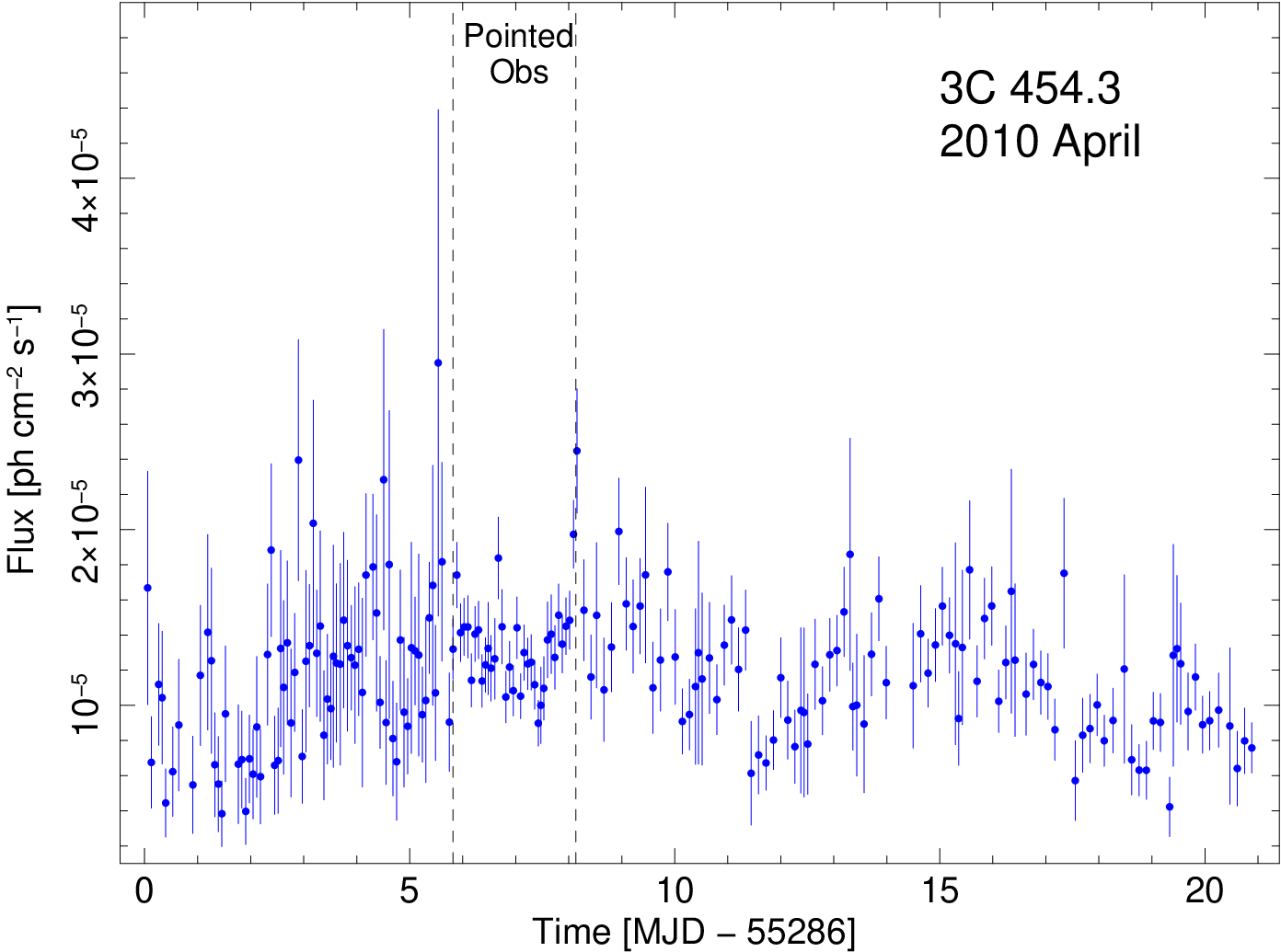}\\
\includegraphics[angle=270,scale=0.35]{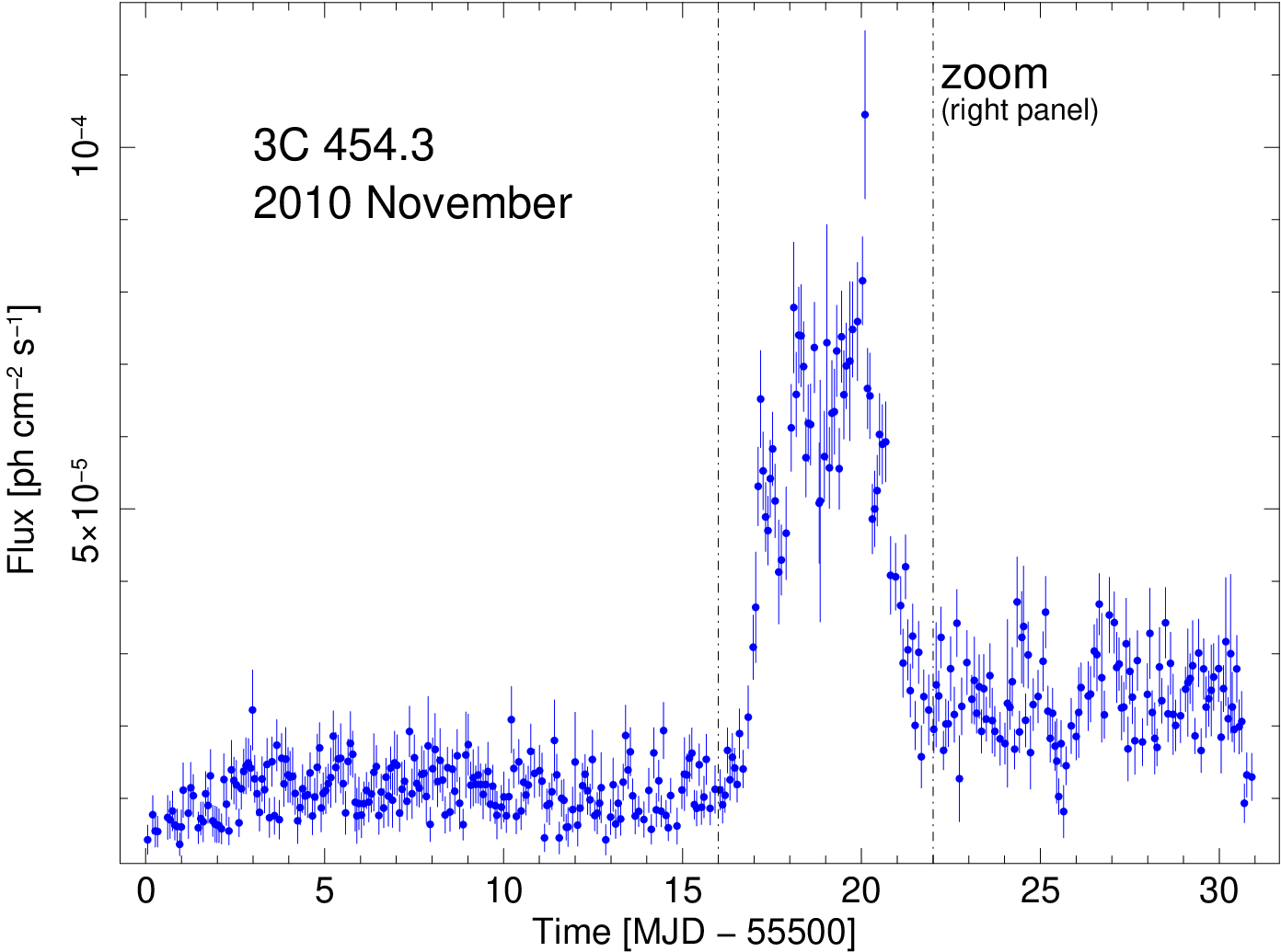}
\includegraphics[angle=270,scale=0.35]{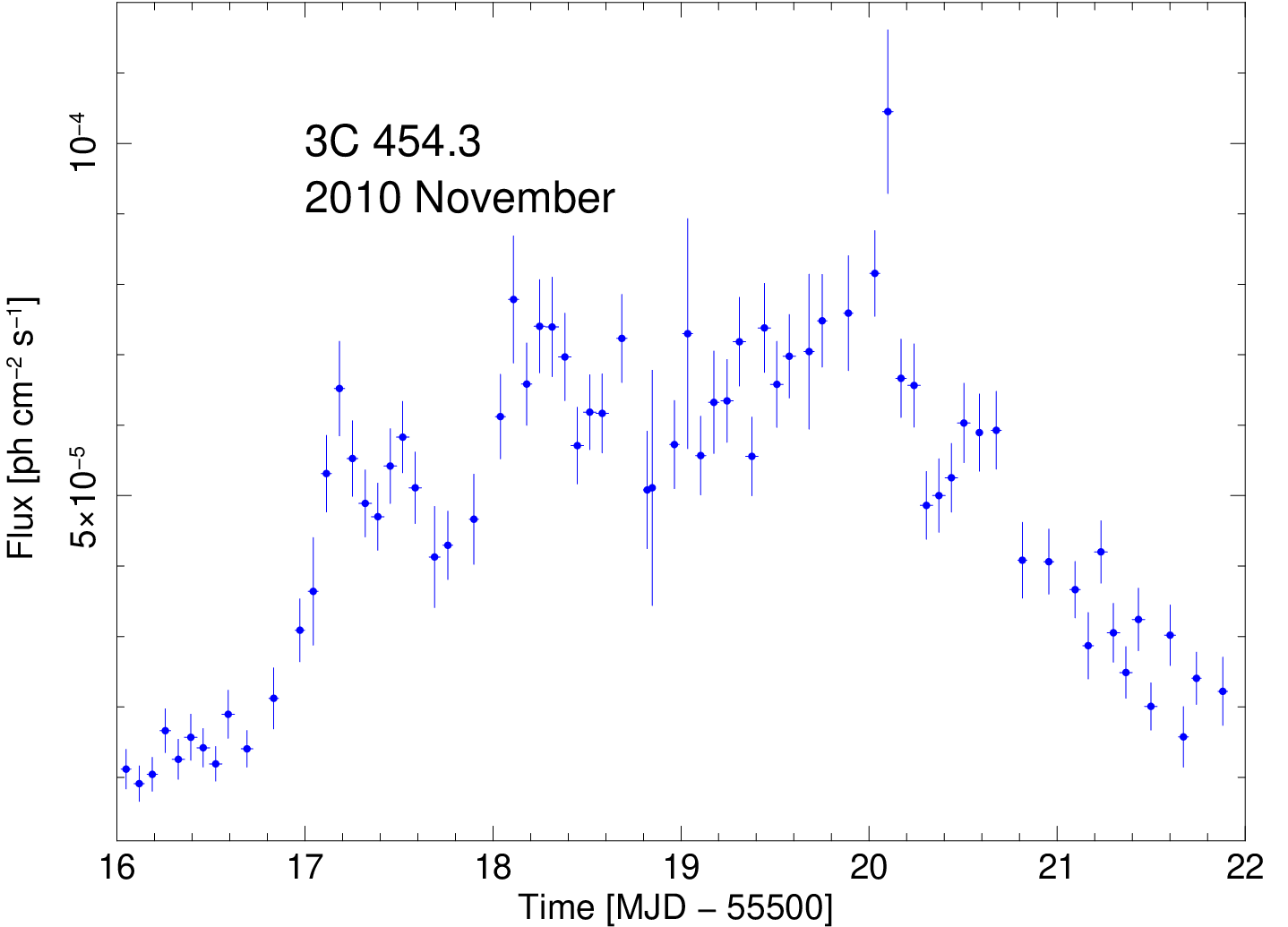}
\caption{Light curves of 3C 454.3 ($E>100$~MeV). (\emph{top left panel}) 2009 December: time starts on 2009 November 30 (MJD 55165), so that the days indicated in abscissa corresponds also to the days of December. (\emph{top right panel}) 2010 April: time starts on 2010 March 31 (MJD 55286), so that the days indicated in abscissa corresponds also to the days of April. The time region between the two dashed lines is that when the pointed observation was done (2010 April 5$-$8). It is clearly evident that the error bars are smaller than those in scanning mode, as a consequence of the better performances of the instrument. (\emph{bottom left panel}) 2010 November: time starts on 2010 October 31 (MJD 55500), so that the days indicated in abscissa corresponds also to the days of November. (\emph{bottom right panel}) 2010 November: zoom of the figure of the \emph{bottom left panel} centered in the period of the highest activity (2010 November 16$-$22), when 3C 454.3 reached the peak of $\sim 10^{-4}$~ph~cm$^{-2}$~s$^{-1}$. Time bins are of the order of a few thousands of seconds and are too small to be visible.}
\label{fig:curva3c454}
\end{figure*}

\begin{table*}
\caption{Summary of characteristic times found. Times $t_0$ and $t$ are in [MJD]; fluxes are in units of [$10^{-5}$~ph~cm$^{-2}$~s$^{-1}$]; the significance of the flux difference is in [$\sigma$]; the absolute values of the observed characteristic time scale $|\tau|$ and the intrinsic one $|\tau_{\rm int}|=|\tau|/(1+z)$ are in hours. The last column indicates a rise (R) or a decay (D) time.}
\begin{center}
\begin{tabular}{llcccccccc}
\hline
Source & Period & $t_0$ & $t$ & $F(t_0)$ & $F(t)$ & Signif. & $|\tau|$ & $|\tau_{\rm int}|$ & R/D\\
\hline
3C~273 & 2009 September & 55094.88299 & 55095.02236 & $0.28\pm 0.14$ & $0.87\pm 0.06$ & 4.2 & $< 4.9$ & $<4.2$ & R\\
\hline
1222$+$216 & 2010 May-June & 55315.76870 & 55315.83500 & $0.82\pm 0.18$ & $0.18\pm 0.07$ & 3.5 & $< 2.3$ & $<1.6$ & D\\
{}            & {}            & 55315.83500 & 55315.90130 & $0.18\pm 0.07$ & $0.80\pm 0.20$ & 3.0 & $< 2.5$ & $<1.7$ & R\\
{}            & {}            & 55322.52962 & 55323.59231 & $0.08\pm 0.04$ & $0.51\pm 0.08$ & 5.4 & $9.5\pm 5.3$ & $6.6\pm 3.7$ & R\\
\hline
3C~454.3 & 2009 December & 55167.16269 & 55167.21717 & $0.85\pm 0.37$ & $2.53\pm 0.56$ & 3.0 & $< 3.4$ & $<1.8$ & R\\ 
{}       & {}            & 55171.33812 & 55171.40897 & $2.06\pm 0.37$ & $0.73\pm 0.23$ & 3.6 & $< 3.6$ & $<1.9$ & D\\
{}       & 2010 November & 55511.48731 & 55511.55361 & $1.32\pm 0.28$ & $0.45\pm 0.22$ & 3.1 & $< 5.0$ & $<2.7$ & D\\
{}       & {}            & 55514.12842 & 55514.19822 & $0.57\pm 0.20$ & $1.63\pm 0.35$ & 3.0 & $< 4.1$ & $<2.2$ & R\\
{}       & {}            & 55514.47072 & 55514.53701 & $1.94\pm 0.39$ & $0.76\pm 0.21$ & 3.0 & $< 4.7$ & $<2.5$ & D\\
{}       & {}            & 55522.66673 & 55522.73602 & $3.42\pm 0.46$ & $1.27\pm 0.59$ & 3.6 & $< 4.3$ & $<2.3$ & D\\
{}       & {}            & 55524.28281 & 55524.34911 & $1.68\pm 0.27$ & $3.71\pm 0.62$ & 3.3 & $< 4.8$ & $<2.6$ & R\\
\hline
\end{tabular}
\end{center}
\label{tab:timescales}
\end{table*}

\section{Data Analysis}
As stated above, we have searched for very high fluxes sources ($F_{\gamma} > 10^{-5}$~ph~cm$^{-2}$~s$^{-1}$ averaged over one day; $E>100$~MeV). We have identified three candidates: 3C 454.3, which has exceeded the flux threshold three times to date (2009 December, 2010 April and November), PKS B1222+216 (2010 May-June) and 3C 273 (2009 September). \emph{Fermi}/LAT data for the above mentioned sources and time periods were downloaded from the \emph{Fermi Science Support Center} at HEASARC\footnote{\texttt{http://fermi.gsfc.nasa.gov/ssc/data/}}. The selected data were screened, filtered, and analyzed as described in greater detail in Foschini et al. (2010), but with a more recent version of the \texttt{LAT Science Tools} (\texttt{v. 9.18.6}) and the corresponding calibration files. 

Moreover, while in Foschini et al. (2010) we could build time bins smaller than the good-time intervals (GTI) because of the better LAT performance during pointed observations (the collected counts are greater than in survey mode by a factor $\sim$~3.5, see also Ackermann et al. 2010), this is no more possible when analyzing LAT data from scanning mode observations, which in turn constitute the majority of the data analyzed in the present work. Searching for the best trade off between a small time bin and the need of significant statistics in each bin, we have found that the best option is to have time bins equal to the GTI, which generally are of the order of a few kiloseconds (a bit less than one orbit). Shorter bins could suffer of artifacts as indicated in the caveats listed in the \texttt{LAT Science Tools} web pages at HEASARC; larger bins would wash out the variability. 

Last, but not least, we have required that the flux in each bin was at least a factor 2 greater than its error, i.e. that there were sufficient events to correctly and significantly reconstruct the source flux. 

The extracted light curves are displayed in Figs.~(\ref{fig:curva3c273}--\ref{fig:curva3c454}). We have noted that 3C~454.3 reached its peak flux of $(1.0\pm0.1)\times 10^{-4}$~ph~cm$^{-2}$~s$^{-1}$ ($E>100$~MeV, corresponding to a luminosity of about $\sim 3\times 10^{50}$~erg~s$^{-1}$) on 2010 November 20 between 01:45 and 03:03 UTC (source ontime $\sim 4.7$~ks). During this time, LAT detected 110 events from the blazar ($22\sigma$ detection). This is the greatest $\gamma$-ray flux ever detected to date from an AGN. It is worth noting that, contrary to previous observations at high-energy $\gamma$ rays, this time 3C~454.3 displayed significant spectral changes in the $\gamma$-ray energy band (Abdo et al. 2011).

Once prepared, we have scanned all the light curves searching for the minimum time of doubling/halving flux:

\begin{equation}
F(t) = F(t_0)\cdot 2^{-(t-t_0)/\tau}
\end{equation}

\noindent where $F(t)$ and $F(t_0)$ are the fluxes at the time $t$ and $t_0$, respectively, and $\tau$ is the characteristic time scale. We have required that the difference in flux is significant at $3\sigma$ level (at least) and we have selected only the results with $|\tau|<5$~hours (including the uncertainties). The results are displayed in Table~\ref{tab:timescales}\footnote{It is worth noting that there are some flares in the curves resolved with a few points. In this case, it would be possible to fit these points and calculate the $\tau$ with smaller errors. For example, in the case of PKS~B1222$+$216 it is possible to measure $\tau=1.5\pm 0.6$ hours (fit of 4 points). However, given all the uncertainties and the low statistics, we prefer to adopt the most conservative approach outlined above.}. 

In the case of 3C~454.3, we note that the less variable period is that of 2010 April, just when the pointed observation was done. This was already noted in Foschini et al. (2010) and Ackermann et al. (2010), although the former reported a $\sim 3\sigma$ change in flux at the end of the pointed observation (in the early hours of 2010 April 8, when there was a short flare; see Fig.~\ref{fig:curva3c454}, \emph{top right panel}) with $\tau \sim 3$~hours, which is not found in the present analysis, because -- as expected -- a greater time bin (1.8~ks in Foschini et al. 2010; 4.8$-$5.7~ks in the present work) smoothed the variability (the measured change in flux is now at $\sim 2.5\sigma$ level). Therefore, since the pointed observation has been already studied in the above cited works and the analysis performed in the present work cannot give better results, because it is best suited for observations in scanning mode, we have decided to drop this period and we report the data in the figures for the sake of completeness.

\section{Discussion}
The masses of the three sources are of the order of some $\times 10^{8}M_{\odot}$ (see Bonnoli et al. 2011, Fan \& Cao 2004, Ghisellini et al. 2010), which means gravitational radii $r_{g}\sim (0.5-1)\times 10^{14}$~cm and a minimum characteristic time $r_{\rm g}/c\sim 0.5-0.9$~hours. The minimum size of the emitting blob of plasma as inferred from the upper limit of variability is $r < c\delta \tau_{\rm int}\sim (20-30)r_{g}$, by considering a typical $\delta = 10$ (Ghisellini et al. 2010). Moreover, the size $r$ of the emitting blob is linked to its distance from the central engine by the relationship $r \sim \psi R$, where $\psi$ is the semi-aperture angle of the jet, generally about $0.10-0.25$ (Ghisellini \& Tavecchio 2009, Dermer et al. 2009). The values of $r$ calculated from the measurements of variability in the present work are $r\sim (1-3)\times 10^{15}$~cm, which then imply $R\sim (1-8)\times 10^{16}$~cm, smaller but comparable with the BLR size, where most of the dissipation occurs in FSRQs (cf Ghisellini et al. 2010). Therefore, although the characteristics time scales reported in the present work are the shortest ever measured in the MeV energy range to date, they are still satisfying the conditions on the spatial dimensions suggested from the standard paradigm.

We can try to understand if these upper limits on $\tau$ could be useful in distinguishing between the location where most of the dissipation occurs. This is particularly important, because during the latest years there was and still is a lively intriguing debate on this topic. On one side, there is a group of researchers supporting the hypothesis that the $\gamma$ rays in the sources powered by the EC process are generated by the interaction with seed photons from the molecular torus (infrared photons), which is placed at some parsecs from the central engine (e.g. Abdo et al. 2010c; Marscher et al. 2008, 2010; Sikora et al. 2009). On the other side, another group of researchers is instead favoring -- with different arguments -- the hypothesis of a dissipation inside the BLR, i.e. less than one tenth of parsec (e.g. Finke \& Dermer 2010, Ghisellini et al. 2010, Poutanen \& Stern 2010, Tavecchio et al. 2010). 

In order to distinguish between the two hypotheses, we have followed the argument on the electrons cooling time as explained in Tavecchio et al. (2010). Basically, the observed cooling time $t_{\rm c}^{\rm obs}$ can be calculated as:

\begin{equation}
t_{\rm c}^{\rm obs} = \frac{3m_{\rm e}c}{4\sigma_{\rm T}U'}\sqrt{\frac{\nu_{\rm seed}^{\rm obs}\Gamma(1+z)}{\nu_{\rm IC}^{\rm obs}\delta}}
\label{eq:cooling}
\end{equation}

\noindent where $m_{\rm e}$ is the electron rest mass, $\sigma_{\rm T}$ is the Thomson cross section, $\Gamma$ is the bulk Lorentz factor, $\nu_{\rm seed}^{\rm obs}$ and $\nu_{\rm IC}^{\rm obs}$ are the peak frequencies of the seed photons ($2\times 10^{15}$~Hz for the broad-line region; $3\times 10^{13}$~Hz for the molecular torus) and of the inverse-Compton ($100$~MeV$=2.4\times 10^{22}$~Hz). $U'$ is the energy density of seed photons (in the comoving frame), which is $\sim 3.8\times 10^{-2}\times \Gamma^2$~erg~cm$^{-3}$ for the broad-line region and $\sim 3.0\times 10^{-4}\times \Gamma^2$~erg~cm$^{-3}$ for the infrared torus (see Ghisellini \& Tavecchio 2009). The values calculated by using Eq.~(\ref{eq:cooling}) have to be compared with the characteristic times observed during the declining phases of the outbursts (see the D values of the last column in the Table~\ref{tab:timescales}). 

By assuming the typical value of $\Gamma \sim \delta \sim 10$ (e.g. Ghisellini et al. 2010), Eq.~(\ref{eq:cooling}) becomes: 

\begin{equation}
t_{\rm c}^{\rm BLR} \sim 0.67\times \sqrt{(1+z)} \, \rm{hours}
\label{eq:coolblr}
\end{equation}

\begin{equation}
t_{\rm c}^{\rm torus}  \sim 10.1\times \sqrt{(1+z)} \, \rm{hours}
\label{eq:cooltorus}
\end{equation}

By substituting the redshifts of the sources, the observed cooling times span from a bit less than one hour in the case of the broad-line region and more than 10~hours for the torus. Therefore, the observed upper limits of a few hours (see $|\tau|$ in Table~\ref{tab:timescales}) of the characteristic time scales favor the BLR (sub-parsec) location of the dissipation zone. 

It would be possible to reconcile the cooling time of EC based on the infrared seed photons from the molecular torus if assuming a large bulk Lorentz factor. For example, if $\Gamma \sim \delta \sim 30$, then Eq.~(\ref{eq:cooltorus}) becomes:

\begin{equation}
t_{\rm c}^{\rm torus}  \sim 1.0\times \sqrt{(1+z)} \, \rm{hours}
\label{eq:cooltorus2}
\end{equation}

However, this is an unlikely hypothesis, in conflict with the values of the jet apparent speed as measured with high-resolution radio observations. Lister et al. (2009), by analyzing a large set of data obtained with high-resolution VLBA observations at 15~GHz in the period 1994$-$2007, have reported that the greatest values of the apparent speed are generally measured at positions very close to the core, at the limit of the angular resolution of the instrument ($\sim 1$~mas). In the cases of the three sources studied in the present work, they reported the values of $\beta_{\rm app}^{\rm max}=21,13,14$ for PKS~B1222$+$216, 3C~273 and 3C~454.3, respectively, which corresponds to Doppler factors roughly in the range $\delta \sim 9-15$ (calculated by assuming $\beta_{\rm app} \sim \sqrt{2\delta\Gamma}$). The locations of the features used to perform these measurements are at $\sim 5-15$~pc, which are the typical distances of the infrared torus. A Doppler factor greater by more than a factor 2 seems to be unlikely, although it cannot be excluded \emph{a priori}: we are studying outstanding high fluxes at $\gamma$ rays and, therefore, we cannot avoid thinking that also radio observations (if any) following these strong outburst could result in exceptional values of $\beta_{\rm app}$. 

\section{Final remarks}
We have searched for the shorter time scale in the high-energy $\gamma$-ray emission from FSRQs. To have the best trade off between the smallest time bins and sufficient statistics for a significant detection, we have analyzed the \emph{Fermi}/LAT data of three FSRQs, whose flux above 100~MeV has exceeded the threshold of $10^{-5}$~ph~cm$^{-2}$~s$^{-1}$. We have found three sources: PKS~B1222$+$216, 3C~273, and 3C~454.3. We were able to set the tightest upper limit on the observed characteristic time scale to date, which were of the order of $<2-3$~hours, depending on the source. This, in turn, suggests that the location of the $\gamma$-ray emission region could be constrained as within the broad-line region, thus disfavoring the other possibility of a dissipation zone beyond the infrared torus. These upper limits are not conclusive yet, because it is still possible to invoke exceptionally great values of the Doppler factor to reconcile the observations with the hypothesis of the dissipation zone being located at parsec scale. 

Incidentally, we have noted that the shortest upper limit measured in this work ($\tau < 2.3$~hours) can be set in the case of PKS~B1222$+$216, which was also recently detected above 100~GeV by MAGIC (70$-$400~GeV, Mariotti et al. 2010, Aleksi\'c et al. 2011) and \emph{Fermi}/LAT (Neronov et al. 2010), the latter by integrating the data collected over several days. The GeV flux resulted to be also extremely variable, with doubling time scale of the order of 10 minutes, which is now the shortest time scale ever detected in a FSRQ (Aleksi\'c et al. 2011). The upper limit estimated in the present work is in agreement with the findings of the MAGIC Collaboration, although the latter work refers to an observation performed on 2010 June 17, at the end of the time period analyzed in the present work (see Fig.~\ref{fig:curva1222}, \emph{left panel}), while the shortest variability measured by \emph{Fermi}/LAT reported here refers to the end of 2010 April, at the beginning of the light curve shown in Fig.~\ref{fig:curva1222}. 

The detection at hundreds of GeV, together with fast flux variability, poses new problems. As known, the BLR in FSRQs is a rich environment of soft photons, which in turn can severely limit the escape of $\gamma$ rays with energies above tens of GeV because of pair production (e.g. Liu \& Bai 2006). Therefore, these photons of hundreds of GeV should come from zones outside the BLR, possibly around the molecular torus as suggested by Tanaka et al. (2011) on the basis of the analysis of $\gamma$-ray spectra, but at such distances from the central black hole, it is expected that the blob -- if linearly expanding with a constant ratio $r/R\sim 0.1-0.25$ -- has reached a size so large to exclude the possibility of changes over hours time scales or less. A possible solution has been recently suggested by Tavecchio et al. (2011), who proposed to explain these observation with something similar to the structured jet proposed for BL Lacs.

\section*{Acknowledgments}
LF thanks C.~D. Dermer for useful discussions on his model. This research has made use of data obtained from the High Energy Astrophysics Science Archive Research Center (HEASARC), provided by NASA's Goddard Space Flight Center. This work has been partially supported by Agenzia Spaziale Italiana (ASI) Grant I/009/10/0.

\end{document}